# Virtual Alignment Method and its application to the dental prostheses and diagnosis

Kyungtaek Jun[1, *], Seokhwan Yoon[2, *], Jae-Hong Lim[3], SeungJoon Noh[4]

[1]Department of Applied Mathematics and Statistics, Stony Brook University, NY 11794-3600, USA

[2]2Department of Prosthodontics & Dental Research Institute, School of Dentistry, Seoul National University, Seoul 03080, South Korea

[3]Pohang Accelerator Laboratory, Pohang University of Science and Technology KR, Gyeongbuk, 37673, South Korea

[4] Weill Cornell Medicine CLC Genomics Resources Core Facility, 1300 York Avenue, New York, NY, 10021, USA

\* Correspondence: ktfriends@gmail.com (K.J.), scott1125@snu.ac.kr (S.Y.)

## Abstract

The recent proposal of a new alignment solution for X-ray tomography, Virtual alignment method (VAM) allowed a more accurate method to remove the possible errors that limit the resolution and clarity of the reconstructed image. In the field of dentistry, the movement of patients during the scanning poses as one of the major factors hindering the final reconstructed image quality. Here, the patient's movement was artificially given to the projection image set and the newly proposed algorithm using the sinogram and the fixed point was applied to the tooth sample to compare the reconstruction image to the actual projection image set. The new alignment method showed promising results by reducing the margin of errors down to a few μm, which will allow the production of high-quality dental prostheses with accuracy and precision. We hope that the newly proposed alignment method can be further investigated to be applied more readily in the field of dentistry to provide better quality images of patients to make a more accurate diagnosis and prostheses.

# Introduction

Cone-beam computed tomography (CBCT) has been a valuable medical imaging technique in the fields of dentistry such as oral surgery, endodontics, and orthodontics. Compared to fan-beam CT, CBCT exposes patients with lower radiation dosage during three-dimensional visualization, and this technique has become a standard for 3D planning of implants[1,2,3,4]. However, CBCT scanning comes with few caveats. Various factors may present visible artifacts in the final reconstructed images such as high-density objects and motion of patients during the examination[4,5]. Due to the nature of CBCT, which requires a lot longer exposure time during the examination, the patient movement presents to be the major obstacle in reducing the artifacts in final reconstructed images. To reduce these limitations, there have been various apparatuses and theories that have been explored to obtain CT image reconstruction from parallel x-ray sources. One technical approach is to convert CBCT into parallel beam computed tomography (PBCT). To achieve this, use of compound refractive x-ray optics and monochromator filters have been investigated thus far[6,7,8]. To acquire clean images in PBCT, one of the most important theories explored was the alignment theory concerning the projection image set. Alignment theory nullifies the effect of patient movement during the CT scanning, which makes it possible to obtain clean cross-sectional images of the body; hence, final reconstructed images with far fewer artifacts.

In the field of computer visualization, many researchers have explored the use of parallel X-ray sources to get ideally aligned projections from misaligned projection set. There have been many attempts to utilize alignment algorithms and image matching techniques to address the translation and tilt error problems[9,10,11,12,13,14] by finding the center of rotation (COR) using a rearranged projection image set[15]. Recently, Jun and Yoon proposed a new alignment solution, which uses fixed points and virtual center of rotation to reconstruct CT images[10]. The newly proposed method was effective in restoring images and provided promising results. Specifically, it eliminated the translation errors in the rotation axis and specimen. Also using a virtual rotation axis eliminated the vertical tilt errors from the rotation axis, which is the tilting in the vertical axis that is orthogonal to the X-ray projection. Furthermore, Jun and Kim have proposed an alignment theory for reconstructing the images of elastic objects in motion using the virtual alignment method (VAM), also called the virtual focusing method[10,16]. The basic concept behind VAM on elastic objects is to treat the elastic motion in the projection image set as if it is a rigid-type sample at each projection set. Then VAM can be used to obtain a clean set of reconstructed slices. Moreover, a virtual multi-alignment theory has also been proposed for reconstructing clean slices of projection image set taken from rigid objects exhibiting multiple motions or elastic objects[17,18].

In this paper, we used artificial molar tooth sample to reproduce the movement patients may exhibit during the CT scan in clinical settings. The patient was not directly involved in the present study, but we manually introduced the errors that may stem from anticipated patient movement during the CT scan. These movements were artificially given to the projection image set and VAM was applied to improve the misalignment. This method was primarily applied to the artificial tooth sample to witness how accurately the ideally aligned reconstructed slices are comparable to the projection image set. In this investigation, we incorporated the patient's movement artificially, which occurs during the CT scan, into the projection image set and presented how VAM can be applied to fix the errors. This method was applied to the tooth sample to compare how accurately the projection set matches with the ideally aligned reconstructed slice. We hope that this result can be further investigated to be applied more readily in the field of dentistry.

# Measurement of sample size from CT Image

**Projection image set acquisition**

X-ray micro-computed tomography (μCT) was performed at beamline 6C Bio Medical Imaging of the Pohang Light Source-II[19]. The beamline uses a multi-pole wiggler as its photon source and provides a monochromatic X-ray beam between 10 keV and 55 keV. For this experiment, we produced a 45-keV X-ray beam illumination using a double crystal monochromator based on silicon (220) reflection (DCM-V2; Vactron, Daegu, South Korea). A specimen was mounted on an air-bearing rotation stage (ABRS-150MP-M-AS, Aerotech, Pittsburgh) that positioned the specimen 36 m from the source. X-ray microscope (Optique, Peter, Lentilly, France) behind the specimen at a distance of 920 mm was comprised of a100-μm-thick CdWO4 scintillator facing (010) direction (Miracrys LLC, Nizhny Novgorod, Russia), an objective lens with a magnification factor of 2 (PLAPON 2X/NA 0.08, Olympus, Tokyo, Japan), and a scientific CMOS camera (Edge 5.5, PCO AG, Kelheim, Germany). The field of view (FOV) was 7.8 mm by 6.6 mm with a pixel size of 3.05 μm. For CT, projections were acquired every 0.5° for a complete 180° rotation. Exposure time was 1.0 s per projection. An additional 5 beam profile images and 5 dark-field images were taken for flat-field correction purposes (Fig. 1). Before CT reconstruction, the projections were rebinned by a factor of four to reduce data size to 1/16, which increased the effective pixel size to 12.2 μm. CT reconstruction was completed using Octopus 8.7 software (XRE, Gent, Belgium) that supported filtered-back projection (FBP) reconstruction.

# Measuring the margins in Computed tomography images

We hypothesized that if the x-ray's light source is a parallel beam, then the projection images of the sample will be projected accurately. With regards to the projection image set obtained using a parallel beam x-ray on rigid samples, alignment solution and multi alignment theory have been found[10,17]. Here, VAM fixes the translation errors from the patient and presents us with an ideally aligned reconstructed slice. Then, the obtained cross-section from the ideally aligned reconstructed slice of tooth sample can be compared with the projection image. To accurately measure the margins in computed tomography images, we first found the region of interest (ROI) inside tooth within the reconstructed slice. Figure 2a shows the axial level 378 of the reconstructed slice. To find the ROI of the cross-section of the tooth, we used Graph Cut from MATLAB to find the rough boundaries of the sample. Then, Active contours and Morphology features were used to highlight regions within the pre-defined boundaries as shown in Fig. 2b in yellow highlights. Similarly, the same method can be used to distinguish the inner areas within the cross-section of tooth sample as shown in Fig. 2c. The segmentations from Figs. 2b and 2c can be combined and applied to the reconstructed slice to obtain the image shown in 2d. Then, each slice in the axial level can be processed in the same manner and stacked to obtain reconstructed volume shown in Fig. 2e.

## Comparison between sizes of samples within reconstructed slice and projection images

We have selected the reconstructed slice at axial level 378 in Fig. 3a to measure the density the cross-section of the tooth in the reconstructed slices acquired from Pohang Accelerator Laboratory. This outline drawing shows the X-ray transmitted through the sample at the projection angle of 0˚. Figure 3b shows the density values corresponding to the height of the tooth from Fig. 3a. As shown in the variations of density at each pixel in Fig. 3b, the boundary surface at which the average density values fall below 80 is surface where x-ray passes through the boundaries of the tooth sample. At the projection angle, we can obtain the back-projected image of the density of sample in Fig. 3c to compare the cross-section of the tooth (Fig. 3d). Also, we compared the sample's cross-section and the back-projection image for all projection angles in Video 1. Figure 3e shows the actual sinogram obtained at axial level 378 of the reconstructed slice. Figure 3f is a sinogram of Fig. 2d using Radon transform. These two sinograms do have some differences in their density values but the marginal patterns are almost exact.

**Comparison of reconstructed volume and projection images of the tooth**

To measure the largest size of the tooth sample in the three-dimensional CT image, we measured the length of the largest pixels within the crown preparation image to calculate the maximum length per unit pixel around the edges of the tooth post crown preparation in the reconstructed volume. We used the region of interest (ROI) as the red rectangle with dotted lines (Fig. 4a). We measured the maximum length of the sample both from reconstructed volume and the projection image set, separately. First, the measurement was done on reconstructed volume. The two points depicted in Fig. 4b has the maximum distance of the segmented area annotated with red rectangles with dotted lines. The image in Fig. 4c shows the combined image of two slices that contain the corresponding points. Two coordinates located from each segmented area were used to measure the maximum of 624.35 pixels. Within the projection image set, we found the maximum length from the projection image corresponding to the projection angle of 167.5° (Fig. 4d). Two coordinates from the projection image were used to measure maximum length of 624.09 pixels. The calculated maximum length from the projection image set and reconstructed volume had an error of less than 1 pixel.

## Results and Discussion

The development of the virtual alignment method for the projection image set obtained using a parallel beam x-ray has allowed achieving CT images with significantly better clarity[10,17]. Using fixed points on the rigid samples, translation errors from the sample, CCD, or stage were adjusted to produce a clean CT image with better resolution. Thus, the segmentation of the ideally aligned CT image with better clarity provides an accurate 3D image of the sample (Fig. 2). The 3D image had a comparable size to the actual sample size within the projection image sets. Despite the inherent errors from digitized images in clearly depicting the boundaries of the samples in a 3D image, the error is still less than half-pixel difference and the error can be reduced by reducing the size of the pixels. The limitation of the pixel sizes comes from the current technological limitations in CT machines. There are continuous efforts put into improving the overall performance of the CT machine and this will allow us to capture an image with smaller pixel sizes; thus, smaller inherent errors from digitized images. The pixel size used in this paper is 12.2 µm. The maximum length across the trimmed tooth sample from the projection image set was 624.35 pixels whereas the measured length from 3D reconstructed volume was 624.09 pixels. This shows that the difference in the measured maximum length of the above two is less than 0.5 pixels. Utilizing a CT machine that can generate images with smaller pixel sizes in the future will minimize the errors and allow us to generate an ideal set of the 3D image.

Conventional CT scans are better suited in capturing images from fixed objects rather than objects with movements. Likewise, CT scans can be utilized in dentistry more affluently compared to the medical field. Fixed prosthesis such as dental crown and bridge only concerns the dental hard tissues and would benefit immediately from the aligned and improved CT images. Also, good quality dental prosthesis requires an accurate impression to be taken from the patient's oral cavity. CAD/CAM work allows up to approximately 100 µm of the margin of errors. However, our methods of alignment reduce the margin of errors down to a few µm and allow the production of high-quality dental prostheses with accuracy and precision. Moreover, we hope that the exceptional quality of dental prosthesis produced will significantly reduce the labor and time spent at the chairside applying the prosthesis to the patients.

Impression taken from the patient using the existing intraoral scanner is known to be more accurate compared to the conventional impression technique. However, the physical size of the scanner is quite big compared to the oral cavity of most patients. Not only patients are put through discomfort due to the size of the scanner but also the saliva and blood within the oral cavity can easily interfere with an accurate acquisition of information. Currently, these limitations have yet to be solved and the margin of error persists to be approximately 30 µm. If the CT scan can be corrected with VAM as suggested, patients can be examined comfortably through a CT scan and the CT image may be able to substitute the need for an impression. Also, the CT scan can be used complementary to the intraoral scanner. Furthermore, CT data that have been corrected through VAM may help in the early finding of the diseases more accurately and precisely.

In this investigation, we have introduced how the VAM utilizes the fixed point to fix the translation errors introduced such as patient's movement, etc. Virtual multi-alignment Method

(VMAM) provides further alignment solutions for rigid objects moving in different directions within a sample. Here, VMAM utilizes number of fixed points throughout the projected image set. These fixed points such as piercing points[16], fiducial markers[20,21,22,23,24,25], center of attenuation[10], and center of high- or low-density spots[10] are all distinguishable from projection images. During the CT scan of a patient's oral cavity, there are several movements to consider: overall movement of the patient's body, independent movement of the maxilla, and independent movement of the mandible. This means that maxilla and mandible cannot be treated as one rigid body or sample. In this case, both maxilla and mandible can be given their fixed points and realigned. This method allows obtaining focused CT images with different translation errors fixed from a single CT examination. We hope to apply this valuable VAM to not only in the field of dentistry but also in the field of medicine.

## Method

The virtual alignment method has been continuously developed since early 2017. The basic concept in VAM is to utilize the rigid sample's fixed points from the projection image set and move the entire projected sample within the projection image set to realign to the user's desire. Then the advanced virtual alignment method was developed for the elastic samples that display a set of motions. Furthermore, the virtual VMAM was developed for samples that have both rigid and elastic characteristics during the scanning. During the CT scanning, maxillary and mandibular can move, which would hinder the resolution obtained from scanning upper and lower tooth. Our goal in this paper is to show the basic theory required to align the tooth utilizing one rigid sample.

In this paper, we focused on explaining the VAM by aligning the misaligned projection image set that has hypothetical random translation errors. Figure 5a shows the sinogram of the sample at axial level 27 when there are translational errors in various directions (up-down, left-right, and front and back) (See Video 2). First, we align the height of the projected sample in the whole projection image set to find the common layer set. To process this alignment, we used the highest point of the sample as the fixed point. The common layer set can be obtained by aligning the entire projection image set using fixed points to eliminate the axial level errors (Fig. 5b and Video 3). Fig. 5c shows the ideally aligned sinogram that has trajectory function of the fixed point $T_{121,-24°}$ (See Video 4). Video 5 shows an ideally aligned projection image set where the fixed point was aligned to the virtual COR $T_{0,\theta}$ and Fig. 5d shows the corresponding sinogram from the common layer 27. Figs. 5e and 5f are reconstructed slice from the sinogram Figs. 5c and 5d, respectively.

# Acknowledgement

The authors would like to acknowledge the help from Pohang Accelerator Laboratory for allowing the usage of Bio Medical Imaging beamline for producing the projection image set, which were used in this manuscript.

# Author contributions Statement

K.J. conceived and designed the theoretical experiments. K.J., S.Y., and J.L. performed the experiments. K.J., and S.N. developed the theoretical solutions. S.Y. conceived and designed the dental samples. J.L. conceived and designed the Bio Medical Imaging beamline. All authors wrote the paper.

# Competing financial interests

The author declares no competing financial and nonfinancial interests.

# Figures

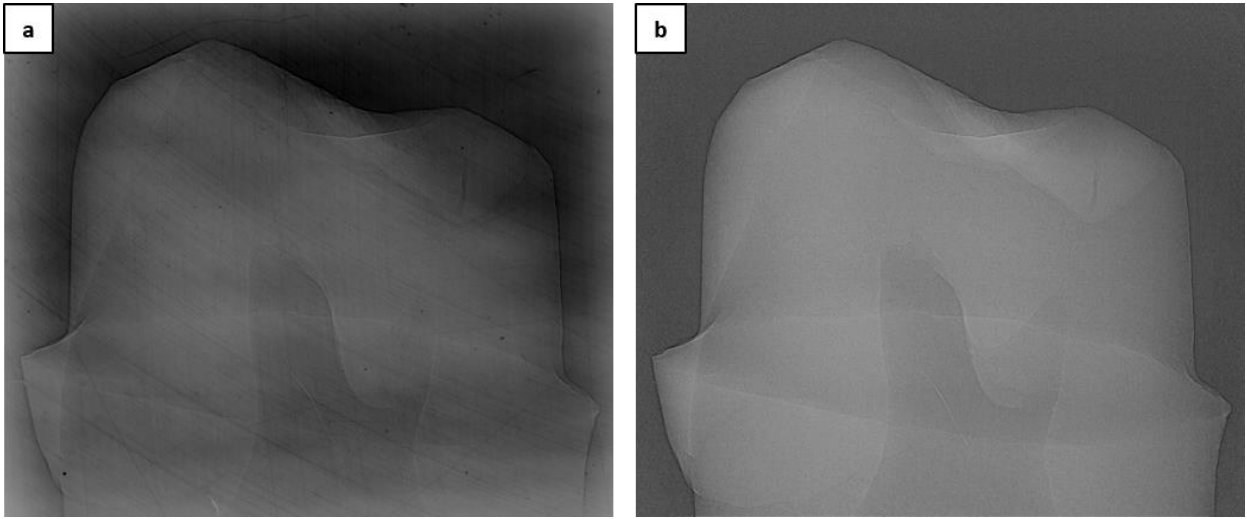

**Figure 1. Normalization of Projection Image.** (a) Projection image acquired from the laboratory. (b) Projection image with flat-field correction applied.

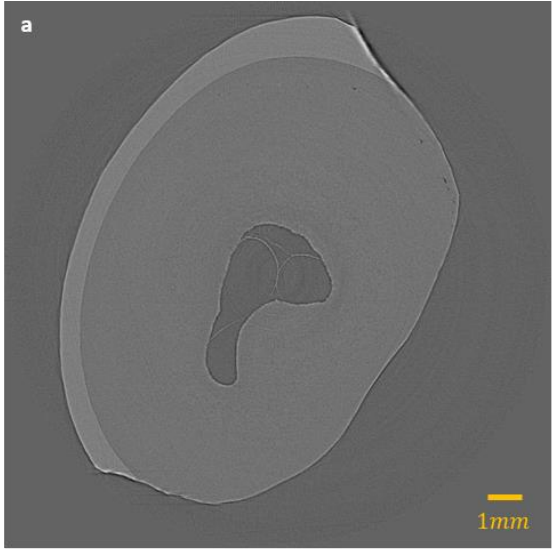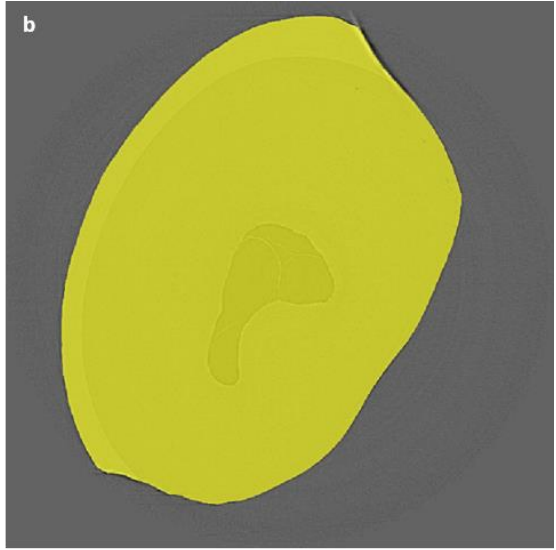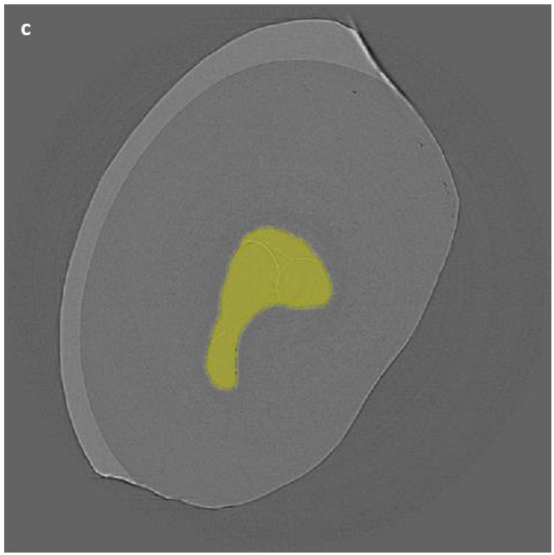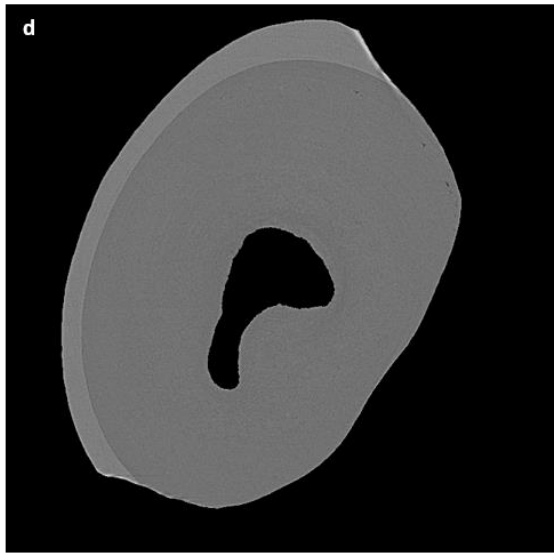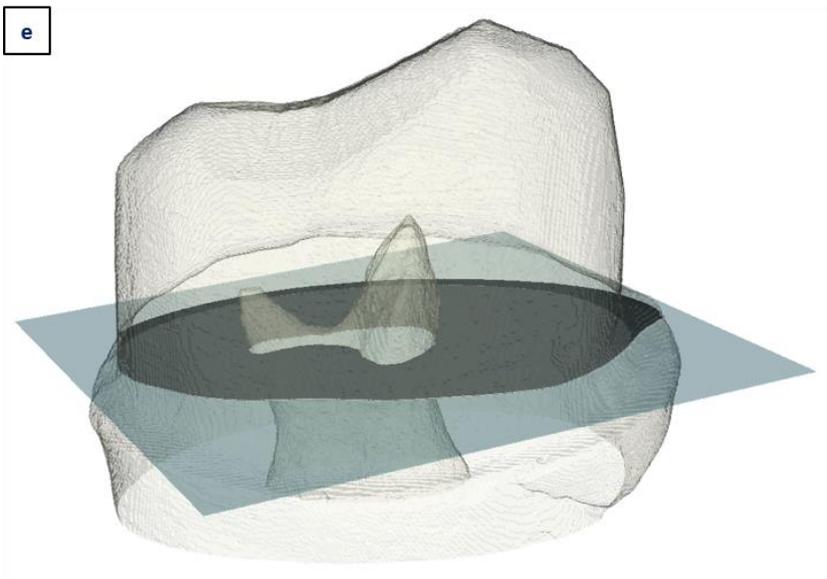

**Figure 2. Calculating region of interest of the reconstructed slice and cross-section of tooth.**

(a) The image is the reconstructed slice of tooth at axial level 378. (b) The image shows the segmentation of the cross-section of tooth and the inner area collectively in yellow. (c) The inner space was distinguished within the cross-section of tooth and segmented in yellow. (d) The image shows the segmented region of interest in the cross-section of tooth applied to the reconstructed slice. (e) Reconstructed volume of a tooth sample and reconstructed slice at axial level 378 has been shown in dark grey colors.

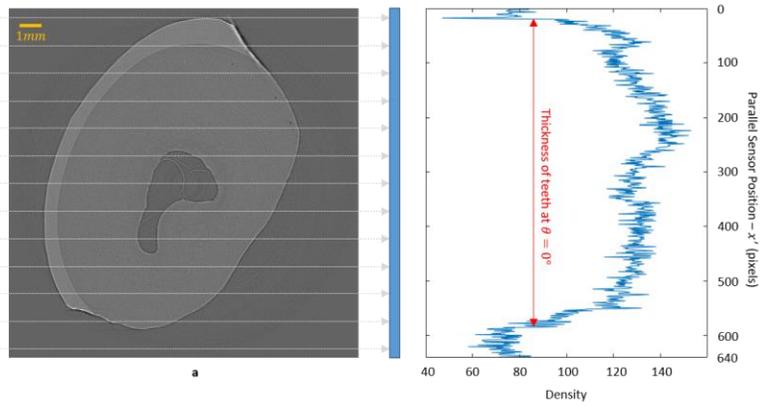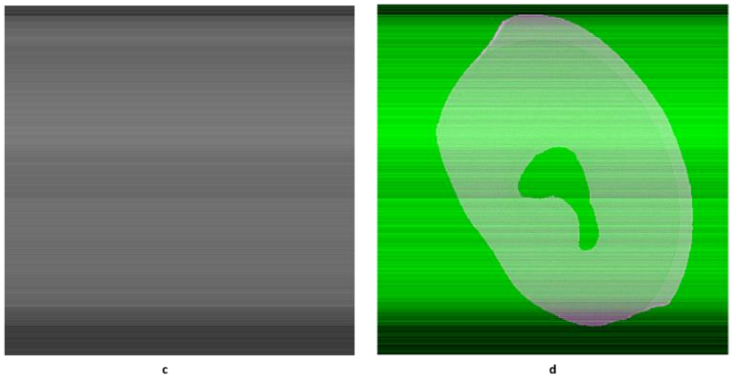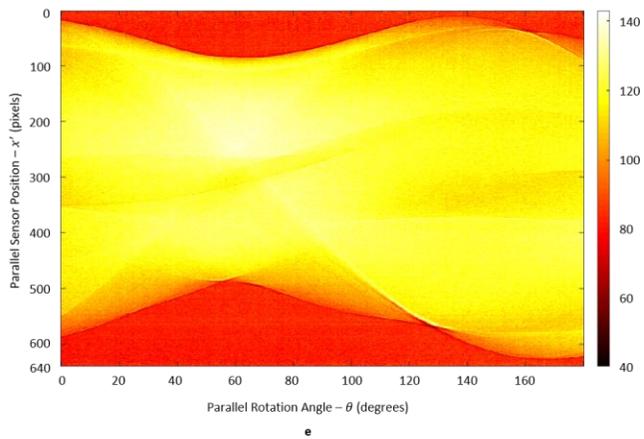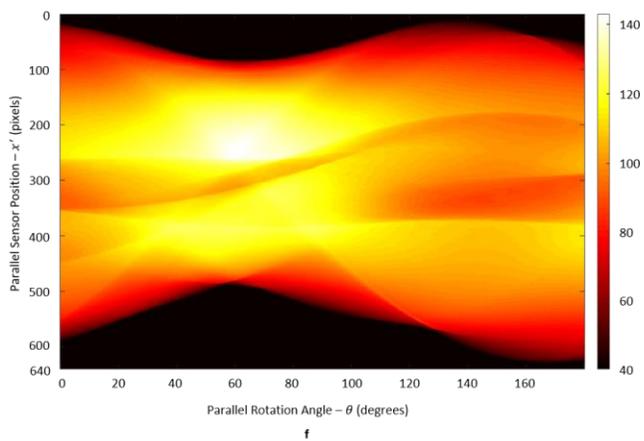

**Figure 3. Measurement of margins around the tooth.** (a) The images show the overview of measured density at CCD when the x-ray transmits through the cross section tooth sample. The gray arrows show the direction of X-ray transmission and blue rectangle shows the outline of a CCD. (b) The image shows the density of the projection image when the X-ray's beams projection pathway aligns with Fig. 3a's dotted arrows. (c) The back-projection image of the density values from Fig. 3b. (d) The image is the comparison of Fig. 2d and 3c. (e) The sinogram is obtained from the reconstructed slice at axial level 378. (f) The sinogram is obtained by using Radon transform on Fig. 2d.

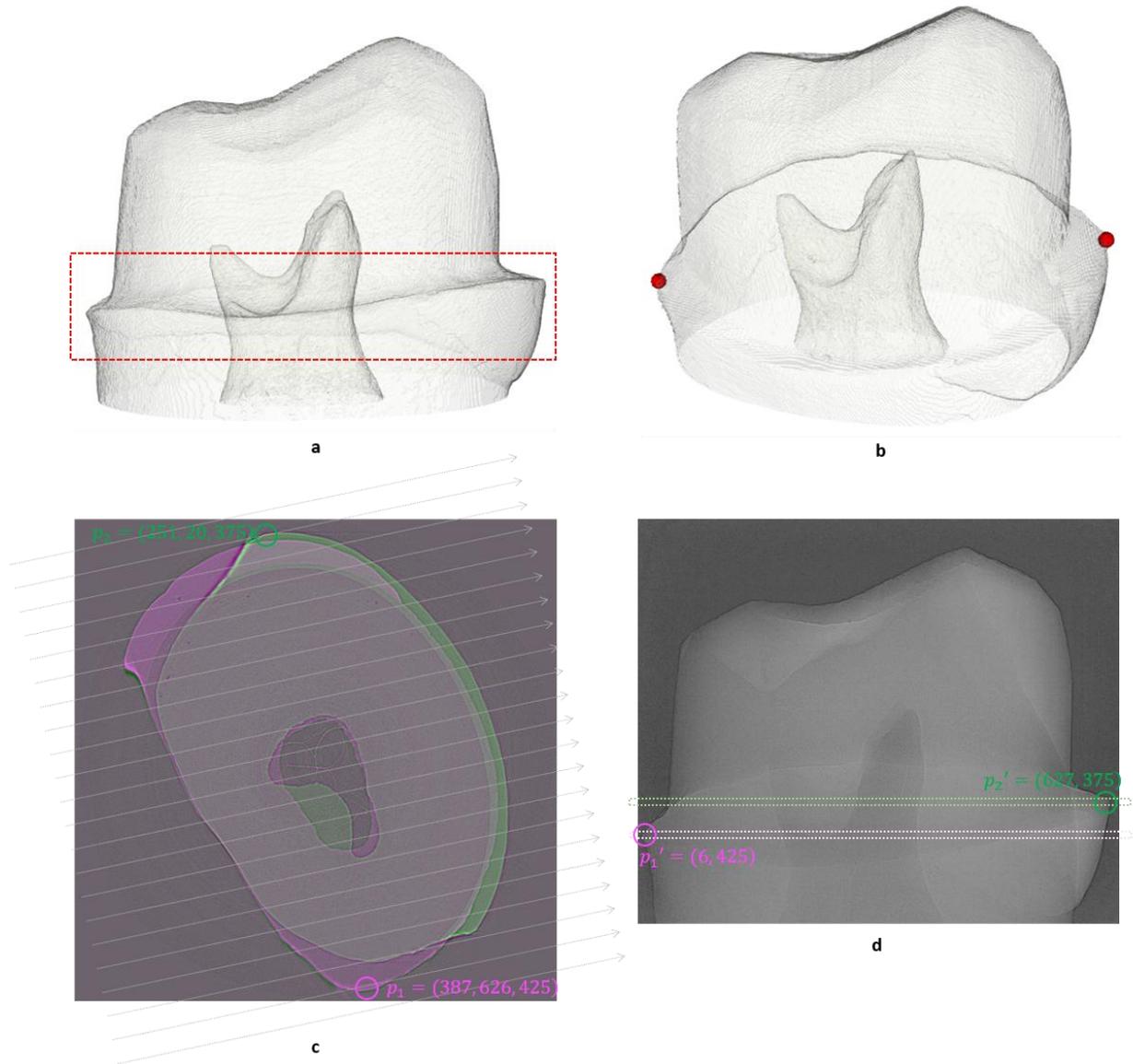

**Figure 4. Measurement of the maximum distance of boundary surface in a reconstructed volume of the crown preparation.** (a) The image is the 3D reconstructed volume of the tooth sample. Red rectangle with dotted lines is the region of interest to configure the maximum distance of boundary surface in crown preparation. (b) The image shows maximum distance in the reconstructed volume. Two red dots in the 3D image shows the two points within the segmented areas from (a) that gives the maximum length. (c) The combined image of two slices contain two points $p_1$ and $p_2$ from Fig. 4b, where first two values indicate the x and y coordinate

for each points, and the number of z-axis indicate the height of the axial level. (d) The projection image at the projection angle of 167.5° has a maximum distance for the projection image set, and this projection image has mostly orthogonal to the line which includes $p_1$ and $p_2$. Two outermost points $p_1'$ and $p_1'$ have the maximum length. Each color for the rectangles corresponds to the matching colors of slices at axial level from Fig. 4c.

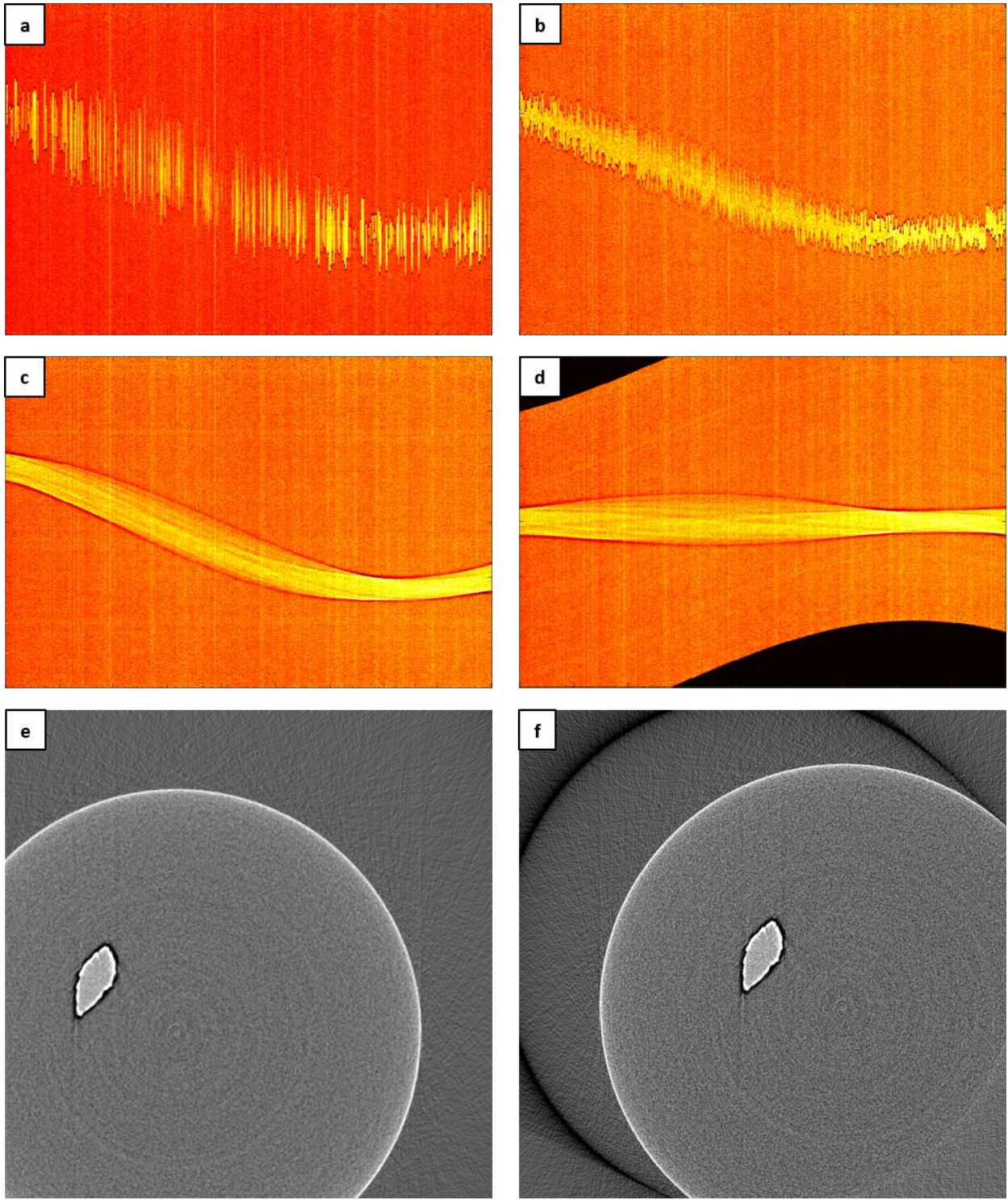

**Figure 5. Illustration of the virtual alignment method applied to the misaligned projection image set.** (a) The sinogram contains the information on the movement of the sample during the scanning in the following directions: up-down, left-right, and front and back. While moving in these directions, this sinogram is from the projection image set of axial level 27. (b) The sinogram shows common layer 27 after fixed points were used to align heights of samples in each slices of the entire projection image set on axial level 27 (c) The ideally aligned sinogram was obtained from applying VAM on sinogram of Fig. 5b. Fixed point from sinogram meets the trajectory function $T_{121,-24°}$. (d) In this sinogram, the fixed point was aligned with the sinogram's center, virtual COR. (e) This is the ideally aligned reconstruction obtained from the sinogram of Fig. 5c. (f) This is the ideally aligned reconstruction obtained from the sinogram of Fig. 5d.